\documentclass{article}%
\usepackage{cite}
\usepackage[left=25mm,right=25mm,top=25mm,bottom=25mm]{geometry}
\usepackage{graphicx}
\usepackage{url}
\usepackage{amsmath,amssymb}
\usepackage{amsmath}
\usepackage{amsfonts}
\usepackage{amssymb}
\usepackage{authblk}
\usepackage{color}%
\providecommand{\U}[1]{\protect\rule{.1in}{.1in}}
\newtheorem{lemma}{Lemma}
\newtheorem{theorem}{Theorem}

\newtheorem{remark}{Remark}
\begin{document}

\title{The Devil is in the Details: Spectrum and Eigenvalue Distribution of the
Discrete Preisach Memory Model}
\author[a]{Tam\'{a}s Kalm\'{a}r-Nagy}
\author[b]{Andreas Amann}
\author[c]{Daniel Kim}
\author[d,*]{Dmitrii Rachinskii}
\affil[a]{Department of Fluid Mechanics, Faculty of Mechanical Engineering, Budapest University of Technology and Economics, Budapest, Hungary; kalmarnagy@ara.bme.hu}
\affil[b]{School of Mathematical Sciences and Tyndall National Institute, University College Cork, Cork, Ireland; a.amann@ucc.ie}
\affil[c]{Texas Academy of Mathematics and Science, Denton, Texas,
USA; danielkim9993@gmail.com} \affil[d]{Department of Mathematical Sciences, The University of Texas at Dallas, USA; dmitry.rachinskiy@utdallas.edu (${}^*$corresponding author)}

\renewcommand\Authands{ and }

\date{}
\maketitle

\begin{abstract}
We consider the adjacency matrix associated with a graph that describes
transitions between $2^{N}$ states of the discrete Preisach memory model. This
matrix can also be associated with the \textquotedblleft
last-in-first-out\textquotedblright\ inventory management rule. We present an
explicit solution for the spectrum by showing that the characteristic
polynomial is the product of Chebyshev polynomials. The eigenvalue
distribution (density of states) is explicitly calculated and is shown to
approach a scaled Devil's staircase. The eigenvectors of the adjacency matrix
are also expressed analytically.
\end{abstract}

\bigskip

	
\section{Introduction}
\label{section_introduction} Hysteresis modeling has been an active area of
research for decades with a wide range of physics-based, phenomenological and
mathematical models
(\cite{visintin1984preisach,brokate1989properties,mayergoyz2003mathematical,krasnosel2012systems,visintin2013differential,science}).
In 1935, F. Preisach proposed his well-known input-state-output model for
ferromagnetic hysteresis \cite{preisach1935magnetische}. The evolution of states in
this model has been later shown to be universal for many important models of
hysteresis with scalar-valued inputs and outputs \cite{brokate2012hysteresis,
krejci} or, more precisely, for all the models which respect Madelung's memory
update rules (also known as hysteresis with return point memory or the wiping
out property \cite{sethna}).

\begin{figure}[tbh]
\begin{center}
\includegraphics[width=0.32\textwidth]{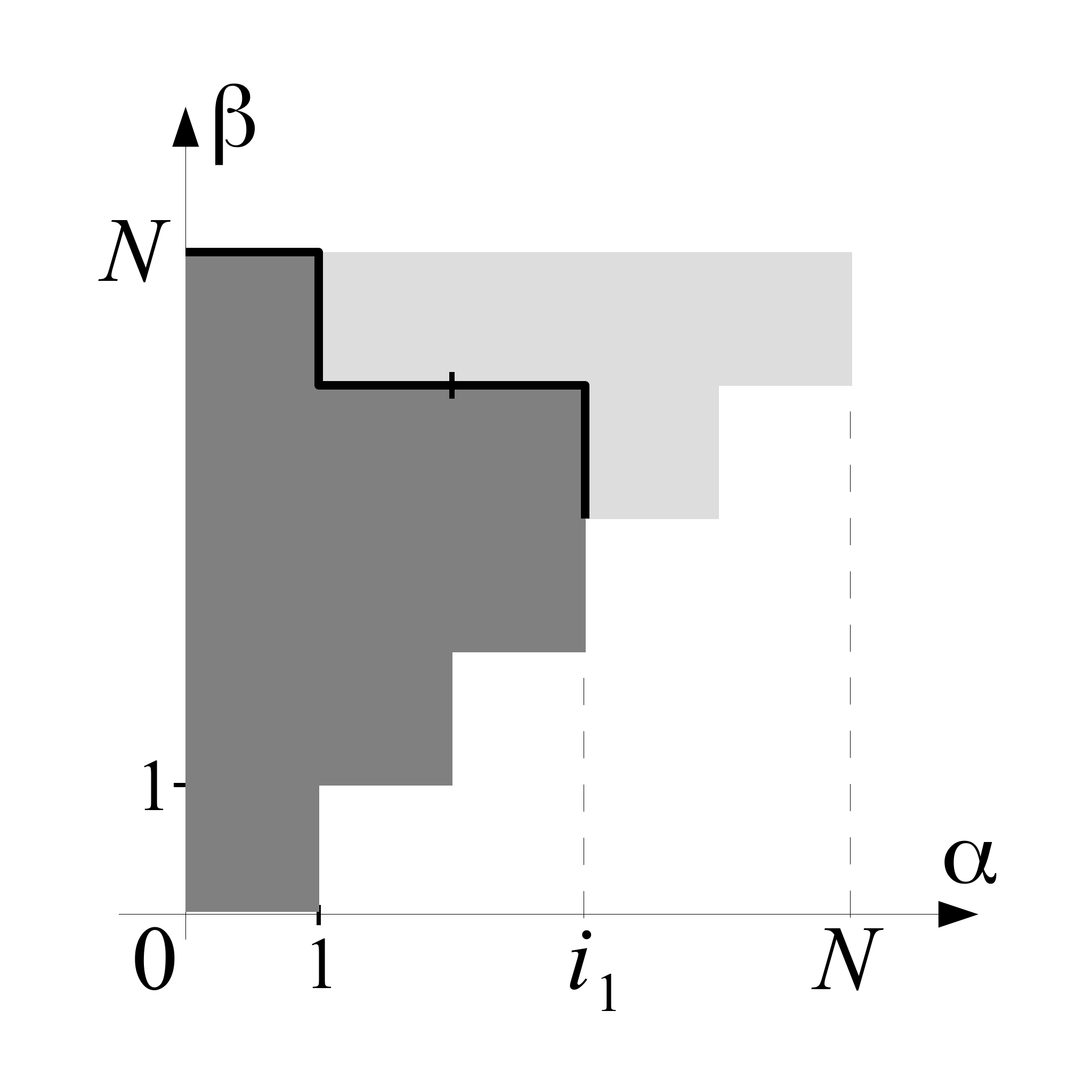}
\ \includegraphics[width=0.32\textwidth]{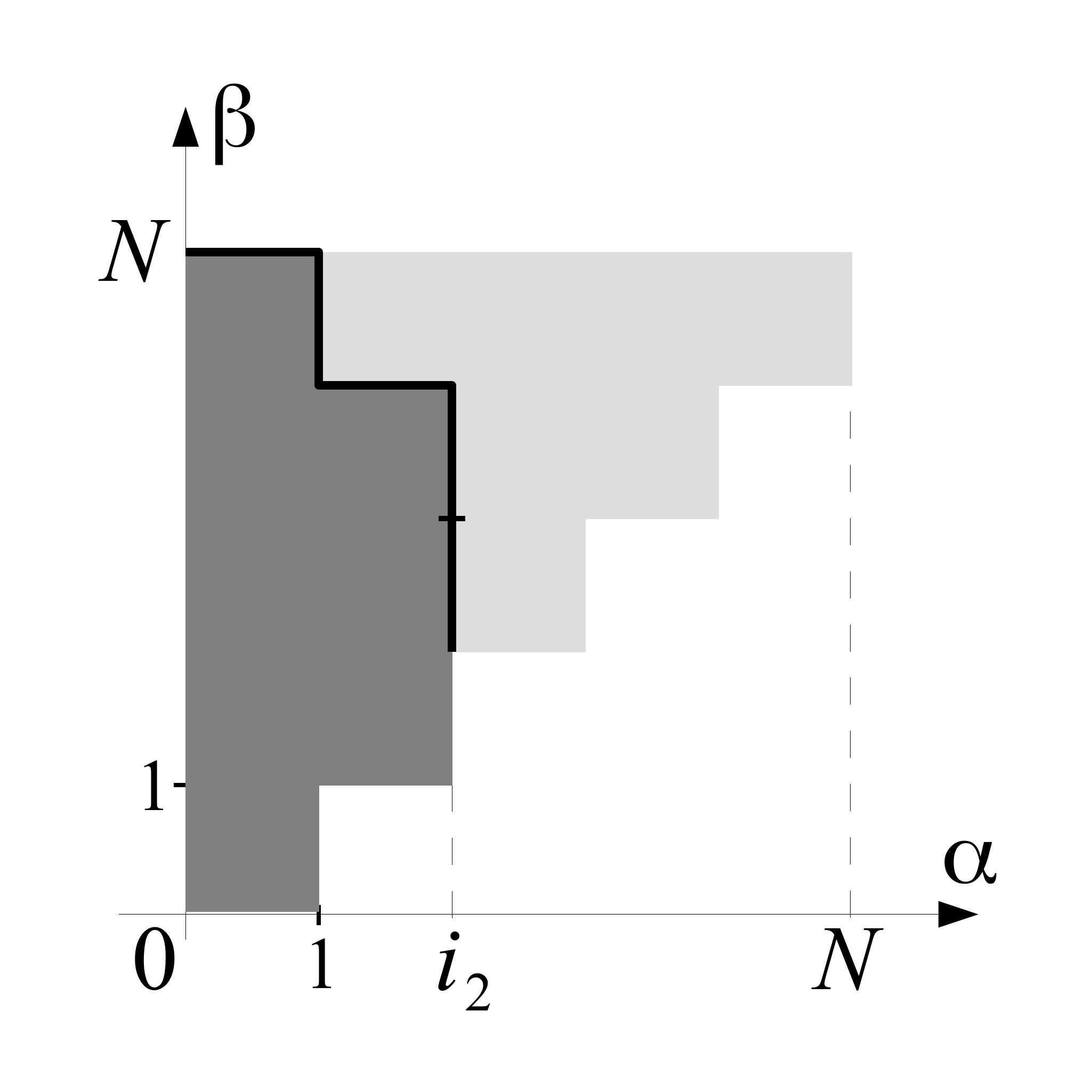}
\ \includegraphics[width=0.32\textwidth]{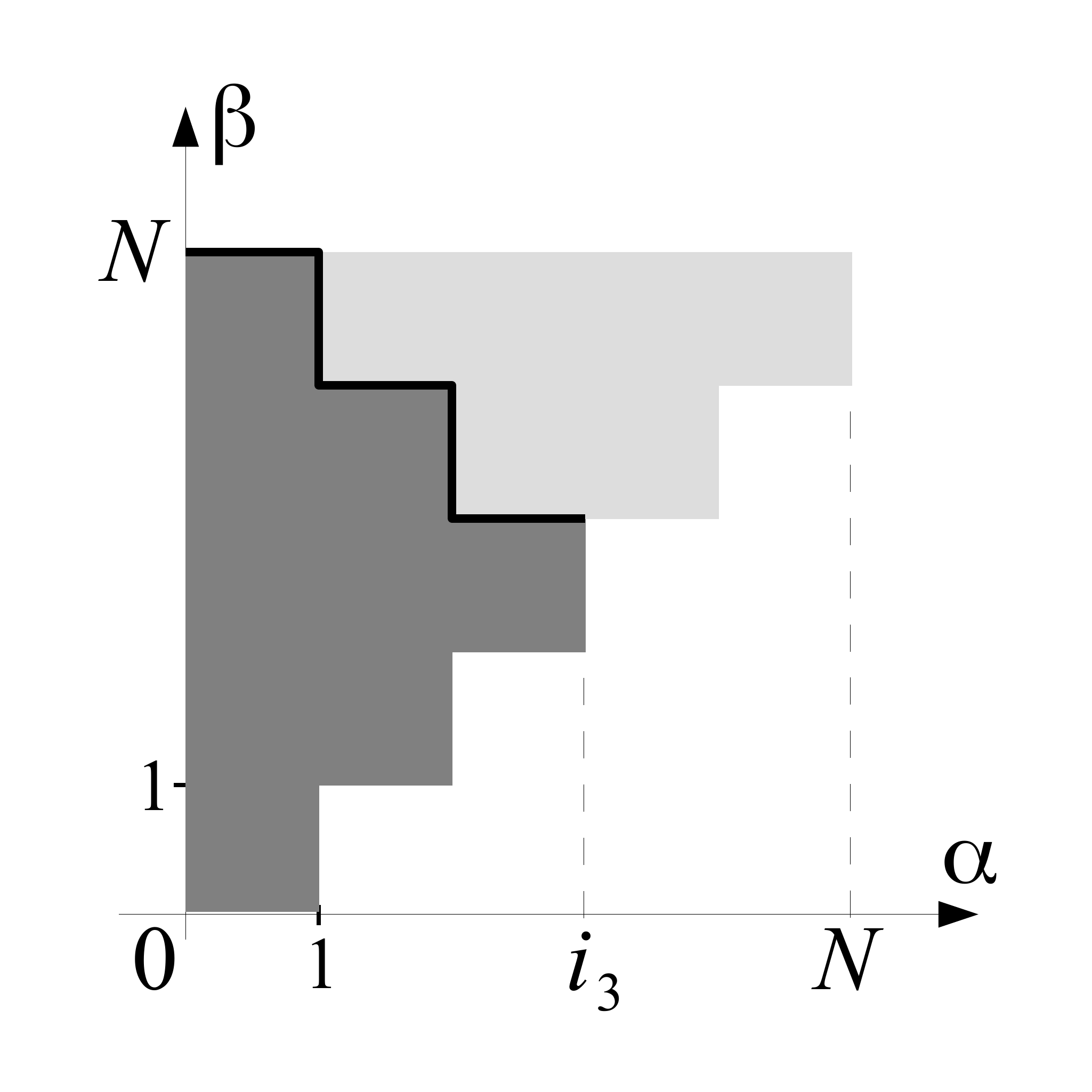}
\phantom{.}\hfil (a) \phantom{mmmmm} \hfil (b) \hfil \phantom{mmmmm} (c)
\hfil \phantom{.}
\end{center}
\caption{The state of the Preisach model is associated with the black polyline
$L$ separating the dark gray and light gray regions. Here $N=5$, and $L$ is
encoded by a 5-tuple of 0s and 1s. The input changes from the value $i_{1}=3$
on panel (a) to the value $i_{2}=i_{1}-1=2$ on panel (b) and back to the value
$i_{3}=i_{2}+1=2$ on panel (c). The state of the Preisach model on these
panels is $(1,0,1,1,0)$, $(1,0,1,0,0)$, and $(1,0,1,0,1)$, respectively,
where $0^\prime$s correspond to vertical and $1^\prime$s correspond to horizontal unit segments.
}%
\label{fig1}%
\end{figure}

In its discrete form, the Preisach model describes the state of magnetic
domains (moments) in a magnetic medium as illustrated in
Figure~\ref{fig1}, where the coordinates $\alpha,\beta$ are parameters (called
thresholds) associated with the magnetic domains. Here, the center of each
dark gray unit box represents a magnetic moment pointing \textquotedblleft
up\textquotedblright\ and the center of each light gray unit box corresponds
to a magnetic moment pointing \textquotedblleft down". The state of the system
is represented by the staircase line $L$ separating the dark and light gray
areas. This line of length $N$ consists of horizontal and vertical unit
segments and it connects the point $(0,N)$ with a point $(i,i)$ on the
diagonal $\alpha=\beta$. Starting at the upper left end $(0,N)$, the line $L$ can be encoded by a unique $N$-tuple of
$0^{\prime}$s and $1^{\prime}$s, corresponding to vertical unit segments and
horizontal unit segments, respectively.

The input $i\in\{0,1,\ldots,N\}$ of the discrete Preisach model (which provides the
coordinates $(i,i)$ of the lower right end of the staircase line $L$)
describes the influence of the external field. The input $i$ can change by
$\pm1$ at each time step, after which the lower right end of $L$ moves accordingly.
Further, when $i$ is increased, all the horizontal segments are passed from
left to right, and when $i$ is decreased, all the vertical segments are passed in
the downward direction. In other words, the rules of updating the state
require that the smallest possible number of unit boxes change their color, see
Figure \ref{fig1}.

A last-in-first-out (LIFO) model can also be associated with these rules.
Consider a one dimensional storage with $N$ spaces. A space is labeled by $0$
if empty and by $1$ if occupied by a box (element). A stored element takes
exactly one space, thus the state of the storage is an $N$-tuple of
$0^{\prime}$s and $1^{\prime}$s. Elements can be added to, or removed from,
the storage through the entrance/exit at the right end. Suppose the following
rules apply: when an element is added to the stock, it is placed to the free
space nearest the entrance; similarly, the element nearest the entrance is
removed when the \textquotedblleft removal\textquotedblright\ operation is
applied. The order in which elements come off the storage according to this
protocol can be characterized as last-in-first-out because we always remove
the most recently added element.\footnote{Note that this protocol is different
from a similar protocol known as stack data type in computer science.} The
transition induced by an increase of the input corresponds to adding an
element to the storage, while the transition induced by a decrease of the
input corresponds to removing an element from the storage.

In the above models, variations of the input induce transitions between the
states, which are equivalent to transitions described by an underlying graph
$\Gamma$. The graph $\Gamma$ is self-similar with interesting properties. The
goal of this paper is to describe and formally prove some of these properties
(for example, we show that the adjacency matrix of $\Gamma$ has a self-similar
eigenvalue distribution), offering an example of analytical treatment of a
self-similar graph to the wider scientific community. Self-similar matrices
and graphs appear in distinct areas. Kostadinov \cite{kostadinov1987fractal}
expresses the free energy of a one-dimensional Ising model with random
couplings in terms of the maximal eigenvalue of a self-similar matrix. He also
relates the calculation for the spectra of molecular type of systems in the
tight binding approximation to the eigenvalues and eigenvectors of
self-similar Hermitian matrices. Stosic et al. \cite{stosic1997residual} find
the analytical expression for the residual entropy of the two-dimensional
Ising model with nearest-neighbor antiferromagnetic coupling in terms of the
fractal Fibonacci matrix. Katsanos and Evangelou \cite{katsanos2001level}
study the level-spacing distribution of a fractal Fibonacci matrix to address
questions related to Anderson localization and quantum chaos. Hsu et al.
introduced Fibonacci cubes as a new class of self-similar graphs
\cite{hsu1993fibonacci}. Ferrand \cite{ferrand2007analogue} considers a
self-similar matrix related to the Thue-Morse sequence. 

Many real networks can also be categorized as self-similar \cite{song2005self}%
. Generating realistic networks is a subject of intense study. Barriere et al.
\cite{barriere2009generalized} investigate the generalized hierarchical
product of graphs. Leskovec et al. \cite{leskovec2010kronecker} propose a
network generation model based on the Kronecker product. Komj\'{a}thy and
Simon \cite{komjathy2011generating} introduce deterministic scale-free
networks derived from a graph directed self-similar fractal. Many books have
been written recently on complex networks and graphs, see, for example
\cite{chung2006complex}.

The eigenvalues of the adjacency matrix are related to important properties of the
graph (for a good introduction to spectral graph theory, see, for example,
\cite{chung1997spectral}). The largest eigenvalue of the adjacency matrix of a
graph plays a key role in several respects, including in synchronization of
oscillators, percolation on directed networks and linear stability of
equilibria of coupled systems.

The paper is structured as follows. In Section \ref{section_preliminaries} we
introduce the necessary notation and define the graphs studied. Section
\ref{section_properties} describes the properties of the adjacency matrix
$A_{n}$. In Section \ref{section_eigenvalues} we look at the eigenvalue
distribution of the (non-symmetric) adjacency matrix. Section \ref{section_eigenvectors}
provides explicit description of the eigenvectors of $A_{n}$. Section
\ref{section_conclusions} concludes our work.

\section{Preliminaries}

\label{section_preliminaries}Consider the set $\{0,1\}^{N}$ of all $N$-tuples
$\mathbf{x}=(x_{0},\ldots,x_{N-1})$ with $x_{i}\in\{0,1\}$ (this is the
Hamming space of all $2^{N}$ binary strings of length $N$). Let us associate a
vertex of a directed graph $\Gamma$ with every $N$-tuple. Define the set of
directed edges of $\Gamma$ according to the rules

\begin{enumerate}
\item[$(H_{1})$] The vertex $\mathbf{0}=(0,0,\ldots0,0,0)$ only has one direct
successor $(0,0,\ldots0,0,1)$;

\item[$(H_{2})$] The vertex $\mathbf{1}=(1,1,\ldots1,1,1)$ only has one direct
successor $(1,1,\ldots1,1,0)$;

\item[$(H_{3})$] Every other vertex $\mathbf{x}$ has two direct successors
$\mathbf{y}$ and $\mathbf{z}$ defined as follows:
\[
x_{i}=0; \quad x_{i+1}=\cdots=x_{N-1}=1 \quad\Rightarrow\quad y_{i}=1,\quad
y_{k}=x_{k} \ \mathrm{for} \ k\ne i;
\]
\[
x_{j}=1; \quad x_{j+1}=\cdots=x_{N-1}=0 \quad\Rightarrow\quad z_{j}=0,\quad
z_{k}=x_{k} \ \mathrm{for} \ k\ne j.
\]

\end{enumerate}

In other words, $N$-tuple $\mathbf{y}$ is obtained by replacing the rightmost
$0$ in the $N$-tuple $\mathbf{x}=(x_{0},\ldots,x_{N-1})$ with $1$ and
$\mathbf{z}$ is obtained by replacing the rightmost $1$ in the $N$-tuple
$\mathbf{x}=(x_{0},\ldots,x_{N-1})$ with $0$. For example, the vertex
$(1,0,1,0,1)$ is connected to the vertices $(1,0,1,0,0)$ and $(1,0,1,1,1)$.

\begin{figure}[tbh]
\begin{center}
\includegraphics[width=0.6\textwidth]{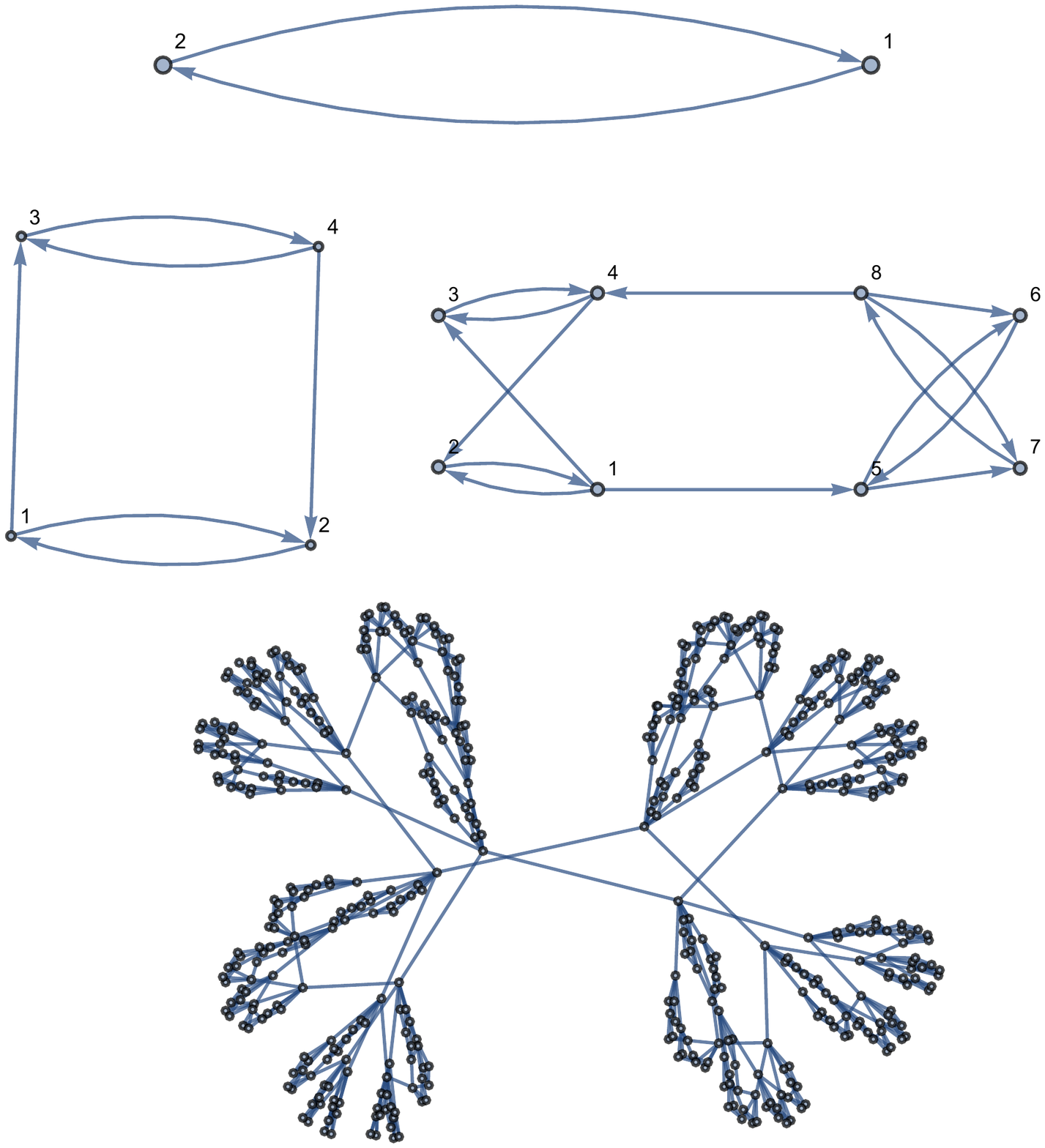}
\end{center}
\vskip-30truemm
\caption{Graphs corresponding to adjacency matrices $A_{1},$ $A_{2},$ $A_{3}$,
and $A_{9}$.}%
\label{fig:graphs}%
\end{figure}

\begin{figure}[tbh]
\begin{center}
\includegraphics[width=0.6\textwidth]{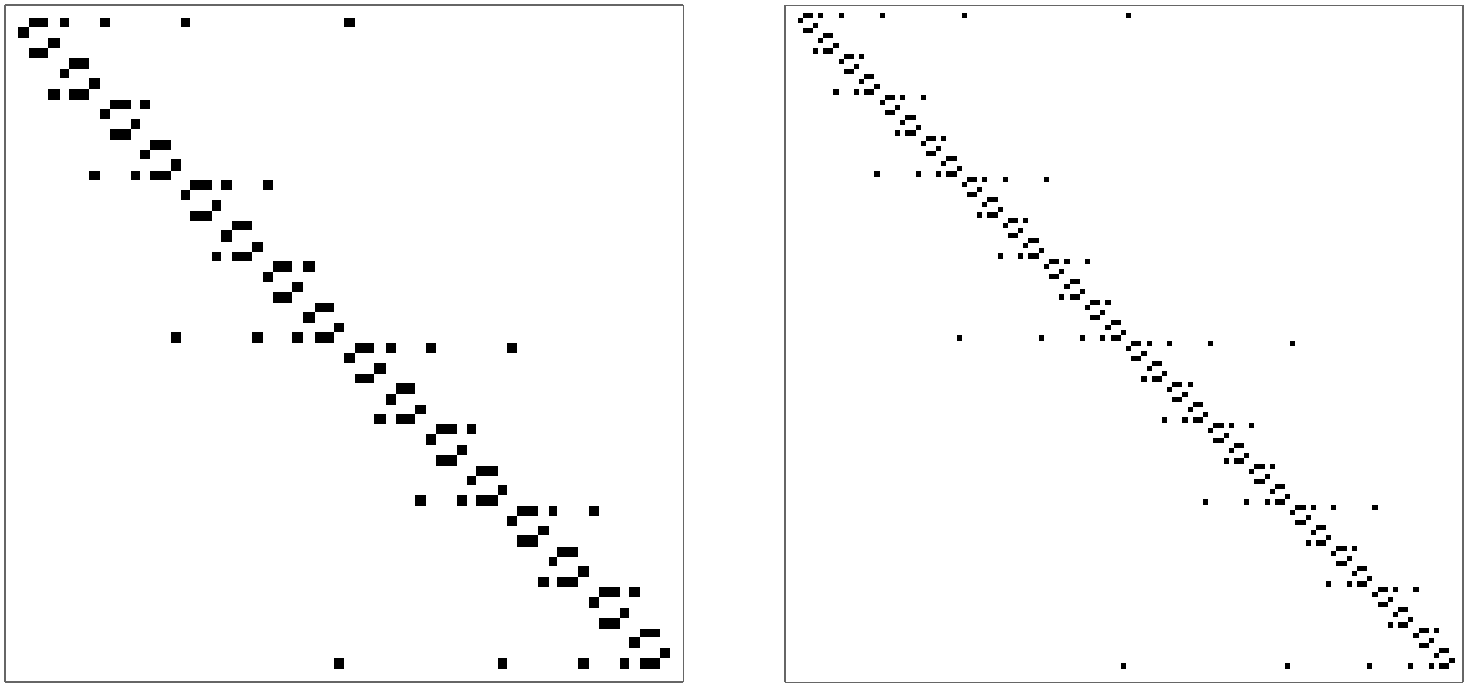}
\end{center}
\caption{Structure plots of $A_{6}$ and $A_{7}$.}%
\label{fig:arrayplots}%
\end{figure}

The adjacency matrix $A_{N}$ (of order $2^{N}$) of graph $\Gamma$ can
recursively be defined by the relations
\begin{equation}
A_{0}=\left(  0\right)  ,\qquad A_{k+1}=\left(
\begin{array}
[c]{c|c}%
A_{k} & J_{k}\\\hline
J_{k}^{\prime} & A_{k}%
\end{array}
\right)  , \label{1}%
\end{equation}
where $J_{k}$ and $J_{k}^{\prime}$ are diagonal matrices of order $2^{k}$
defined by
\begin{equation}
J_{k}=\mathrm{diag}\,\{1,0,0,\ldots,0,0\},\qquad J_{k}^{\prime}=\mathrm{diag}%
\,\{0,0,\ldots,0,0,1\}. \label{J}%
\end{equation}
For example,
\[
A_{1}=\left(
\begin{array}
[c]{cc}%
0 & 1\\
1 & 0
\end{array}
\right)  ,\qquad A_{2}=\left(
\begin{array}
[c]{cc|cc}%
0 & 1 & 1 & 0\\
1 & 0 & 0 & 0\\\hline
0 & 0 & 0 & 1\\
0 & 1 & 1 & 0
\end{array}
\right)  .
\]
Due to its recursive definition, $A_{n}$ is a block-hierarchical matrix, i.e.
a matrix where hierarchically nested growing blocks are placed along the
diagonal, where each (sub-)block is again a block-hierarchical matrix itself
\cite{gutkin2011spectral}. Figure \ref{fig:graphs} illustrates the graphs
corresponding to $A_{1},A_{2},$ $A_{3}$, and $A_{9}$, while Figure
\ref{fig:arrayplots} demonstrates the connection structure of the adjacency
matrices $A_{6}$ and $A_{7}$ (the 1's are shown as black dots).



\begin{remark}
Matrix $\frac{1}{2}A_{N}$ can be viewed as the transition matrix of a discrete
time stochastic process naturally associated with the graph $\Gamma$. This
process transits with equal probability $1/2$ from any vertex (state)
$\mathbf{x}\neq\mathbf{0},\mathbf{1}$ to either of its two direct successors
along the directed edges. If the process is either in state $\mathbf{0}$ or
$\mathbf{1}$, then it transits to the direct successor of this state with
probability $1/2$ or terminates with probability $1/2$. Since the spectral
radius $\rho(\frac{1}{2}A_{N})$ is less than $1$, this process almost surely
terminates in finite time. However, the leading eigenvalue of the matrix
$\frac{1}{2}A_{N}$ tends to $1$ as $N\rightarrow\infty$. Hence, the mean
termination time tends to infinity in this limit.
\end{remark}

Let us denote by $\Gamma^{\prime}$ the graph obtained from $\Gamma$ by adding
self-loops at vertices $\mathbf{0}$ and $\mathbf{1}$. This is achieved by
replacing rules $(H_{1})$, $(H_{2})$ with

\begin{itemize}
\item[$(H_{1}^{\prime})$] The vertex $\mathbf{0}=(0,0,\ldots0,0,0)$ has two
direct successors, $(0,0,\ldots0,0,1)$ and $\mathbf{0}$;

\item[$(H_{2}^{\prime})$] The vertex $\mathbf{1}=(1,1,\ldots1,1,1)$ has two
direct successors, $(1,1,\ldots1,1,0)$ and $\mathbf{1}$.
\end{itemize}

\noindent Let $A_{N}^{\prime}$ denote the adjacency matrix of the graph
$\Gamma^{\prime}$. Then, $\frac{1}{2}A_{N}^{\prime}$ is the transition matrix
of the stochastic process associated with the graph $\Gamma^{\prime}$, which
transits from any vertex to either of its two directs successors with equal
probability $1/2$. This is a Markov chain because $A_{N}^{\prime}$ is a
stochastic matrix. The unique stationary probability distribution for this
Markov chain has been described in \cite{amann2012characterization}, where the
Preisach model with random input was considered\footnote{One can also consider
continuous time Markov chains associated with the graphs $\Gamma$ and
$\Gamma^{\prime}$ with the transition rate matrices $A_{N}-\mathbb{I}_{2^{N}}$
and $A_{N}^{\prime}-\mathbb{I}_{2^{N}}$, respectively, where $\mathbb{I}_{n}$
denotes the identity matrix of order $n$.}. Properties of the stochastic output of the Preisach model under
various random inputs have been characterized in \cite{may,dim,radons2008spectral,siam,radons1,arxiv} (in the context of this work, the output is the area (measure) of the dark gray region in Fig.~\ref{fig1}).

\section{Properties of $A_{n}$}

\label{section_properties}
\begin{figure}[tbh]
\begin{center}
\includegraphics[width=0.62\textwidth]{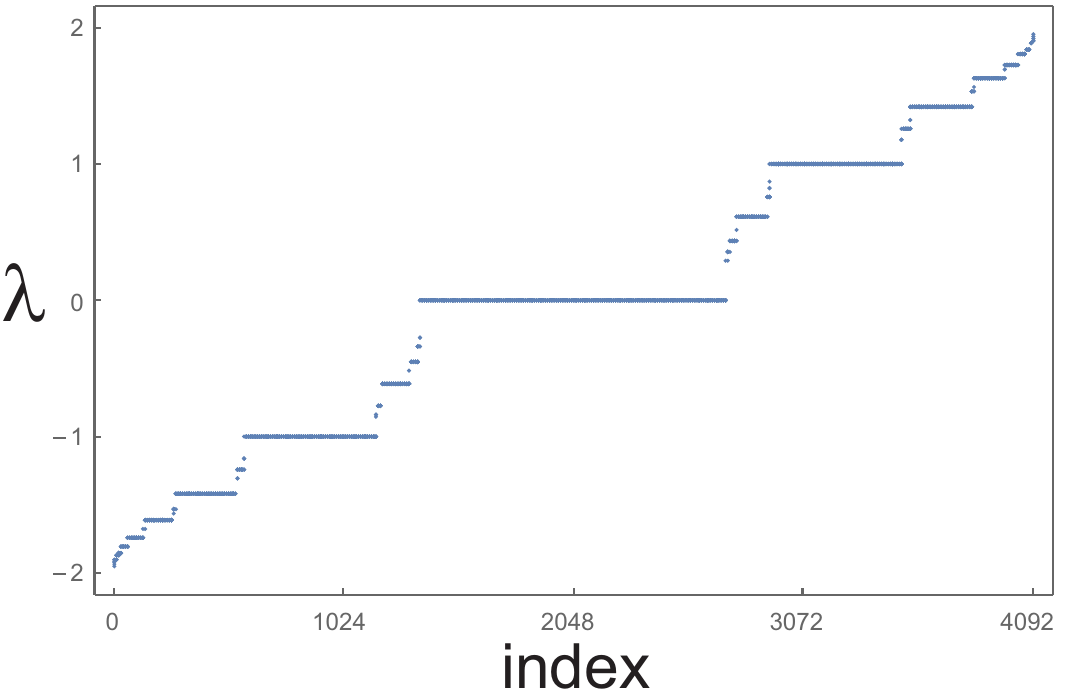}
\end{center}
\caption{Eigenvalue distribution of the matrix $A_{12}$.}%
\label{Fig:eigenvaluedistribution}%
\end{figure}

Spectral graph theory (see, for example,
\cite{chung1997spectral}) relates the eigenvalues of the adjacency matrix to
other properties of the graph. We are thus interested in the spectrum%

\begin{equation}
\mathrm{sp\,}\left(  \frac{1}{2}A_{N}\right)  =\{\lambda\in\mathbb{C}%
:\det(\frac{1}{2}A_{N}-\lambda I)=0\},
\end{equation}
and eigenvalue distribution%
\begin{equation}
F_{N}(x)=\frac{\#\left\{  \lambda\in\mathrm{sp\,}\left(  \frac{1}{2}%
A_{N}\right)  :\lambda<x\right\}  }{2^{N}},
\end{equation}
of the matrix $\frac{1}{2}A_{N}$ and the $N\rightarrow\infty$ limit of these
objects. Here $\#\Omega$ denotes the cardinality of the set $\Omega$.
Figure~\ref{Fig:eigenvaluedistribution} shows the Devil's staircase eigenvalue
distribution for $A_{12}$.

\begin{figure}[tbh]
\begin{center}
\includegraphics[width=0.6\textwidth]{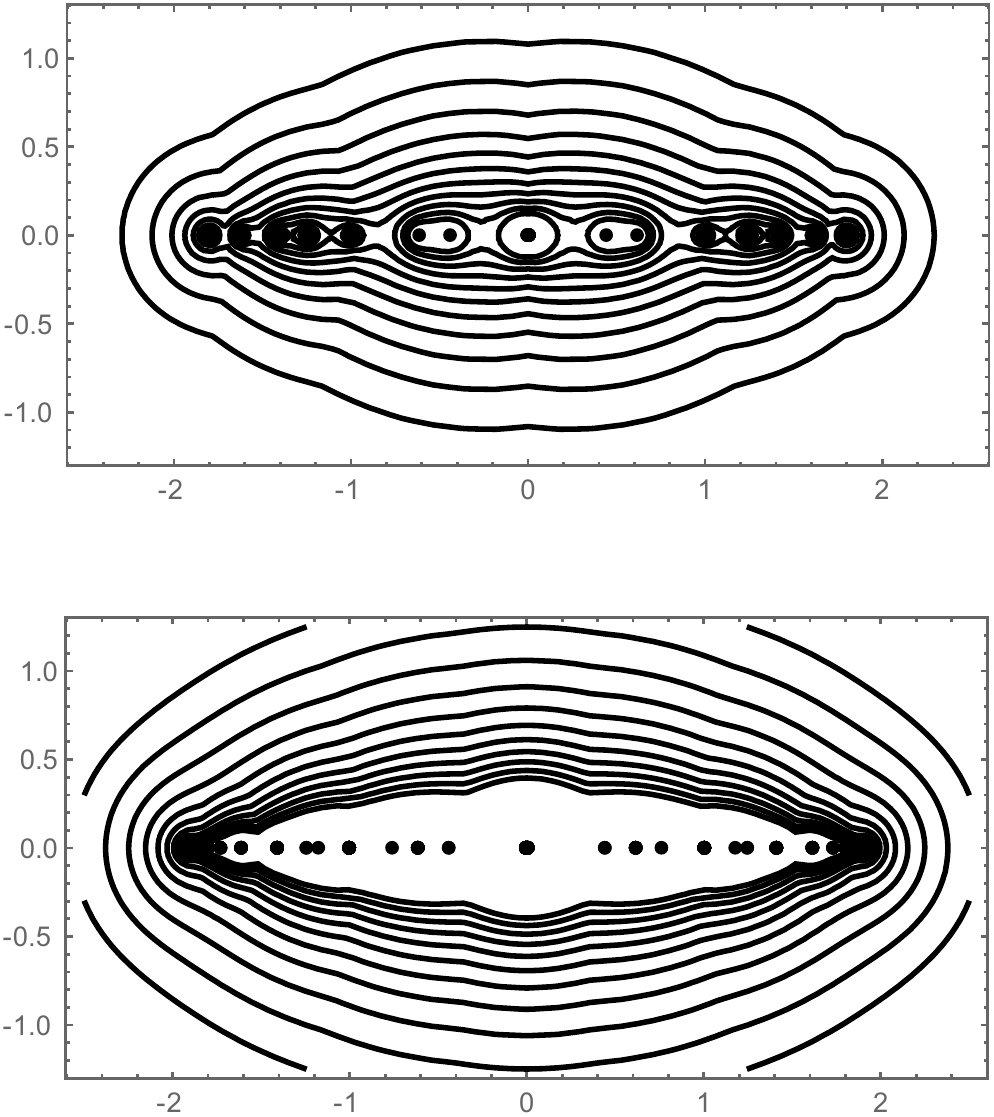}
\end{center}
\caption{Pseudospectra of $A_{5}$ and $A_{8}$.}%
\label{fig:pseudospectra}%
\end{figure}

Xiuqing and Youyan \cite{xiuqing1992spectral} and Fu et al.
\cite{fu1997perfect} studied one-dimensional Fibonacci-class quasilattices and
showed that for these types of lattices the energy spectrum exhibits a
staircase behavior. He et al. \cite{he2000laplacian, he2003trees} studied a
family of trees (the interior nodes have degree $k$ and the boundary nodes
have degree 1) and found that the eigenvalue distributions approach a
piecewise constant \textquotedblleft Cantor function.\textquotedblright
Spectral properties of structured matrices have been extensively studied.
Banded Toeplitz matrices are the topic of the book by Bottcher and Grudsky
\cite{bottcher2005spectral}. A classical result of Schmidt-Spitzer
\cite{schmidt1960toeplitz} and Hirschman \cite{hirschman1967spectra} is that
the eigenvalues accumulate on a special curve in the complex plane and the
normalized eigenvalue counting measure converges weakly to a measure on this
curve as $N\rightarrow\infty$. Duits and Kuijlaars \cite{duits2008equilibrium}
study the limiting eigenvalue distribution of $N\times N$ banded Toeplitz
matrices as $N\rightarrow\infty$, and characterize the limiting measure in
terms of an equilibrium problem. Even though the spectrum of the limiting
operator need not mimic that of the finite-dimensional operator, pseudospectra
and numerical ranges behave nicely \cite{bottcher2005spectral}. The
pseudospectra of $A_{5}$ and $A_{8}$ are shown in
Figure~\ref{fig:pseudospectra}.

Interestingly, Devil's staircases have been associated with the one-dimensional Ising model with antiferromagnetic interactions \cite{aubry}. Exact eigenvalues and eigenvectors of different one-dimensional quantum many-body models starting with the Hamiltonian of the antiferromagnetic Heisenberg model have been obtained using the Bethe ansatz \cite{bax,and}.

\subsection{Spectrum of the adjacency matrix}

\label{section_spectrum}

Denote by $U_{k}(\lambda)$ the Chebyshev polynomial of the second kind of
degree $k$ \cite{mason2002chebyshev}. For example, they can be defined by the
recursive relations
\begin{equation}
{\displaystyle{\begin{aligned}U_{0}(\lambda)&=1; \quad U_{1}(\lambda)=2\lambda;\quad U_{k+1}(\lambda)=2\lambda\,U_{k}(\lambda)-U_{k-1}(\lambda),\end{aligned}}%
} \label{urecursive}%
\end{equation}
or by the explicit formula%
\[
U_{k}(\cos(\vartheta))={\frac{\sin((k+1)\vartheta)}{\sin\vartheta}}.
\]
The zeros of $U_{k}(\lambda)$ are given by
\begin{equation}
{\displaystyle\lambda_{i}=\cos\left(  \pi{\frac{i+1}{k+1}}\right)  },\qquad
i=0,\ldots,k-1. \label{zeros}%
\end{equation}

The main result of this paper is the following theorem.

\begin{theorem}
\label{t1} The characteristic polynomial of the adjacency matrix $A_{N}$ of
the graph $\Gamma$ defined by $(H_{1})$, $(H_{2})$, $(H_{3})$ equals
\begin{equation}
\chi_{N}(\lambda)=U_{N+1}(-\lambda/2)\prod_{i=0}^{N-1}\bigl(U_{i}%
(-\lambda/2)\bigr)^{2^{N-i-1}},\qquad N\geq1. \label{pol}%
\end{equation}

\end{theorem}

Formulas \eqref{zeros} and \eqref{pol} explicitly define the spectrum of the
matrix $A_{N}$.

\begin{theorem}
\label{t2} The characteristic polynomial of the adjacency matrix
$A_{N}^{\prime}$ of the graph $\Gamma^{\prime}$ defined by $(H_{1}^{\prime})$,
$(H_{2}^{\prime})$, $(H_{3})$ equals
\begin{equation}
\label{pol1}\chi_{N}^{\prime}(\lambda)=(2-\lambda)U_{N}(-\lambda/2)\prod
_{i=0}^{N-1} \bigl(U_{i}(-\lambda/2)\bigr)^{2^{N-i-1}}, \qquad N\ge1.
\end{equation}

\end{theorem}

Comparing formulas \eqref{pol} and \eqref{pol1}, one can see that adding the
two self-loops to the graph $\Gamma$ at vertices $\mathbf{0}$ and $\mathbf{1}$
changes exactly $N+1$ eigenvalues in the spectrum of the adjacency matrix;
namely, the roots of the highest degree Chebyshev polynomial $U_{N+1}$ are
replaced with the roots of the Chebyshev polynomial $U_{N}$ and the leading
eigenvalue $1$.

The proof of Theorem \ref{t1} is presented in the next section. The proof of
Theorem \ref{t2} is similar and is omitted.

The appearance of Chebyshev polynomials in the characteristic function of
$A_{n}$ is not entirely surprising: characteristic functions of tri-diagonal
and other structured matrices involve Chebyshev polynomials. Tri-diagonal
matrices are naturally associated with a one-dimensional walk on a path graph.
Our graph has vertices of out-degree of $2$ and thus corresponds to a
one-dimensional walk with longer-range jumps.

\subsection{Proof of Theorem \ref{t1}}

\label{section_proof1}Consider the matrix $T_{k}=A_{k}-\lambda\,\mathbb{I}%
_{2^{k}}$,
 where $\mathbb{I}_{n}$ denotes the identity matrix of order $n$.
With this notation, the characteristic polynomial $\chi_{k}=\chi_{k}(\lambda)$
of $A_{k}$ equals
\[
\chi_{k}=\mathrm{det}\,T_{k}.
\]
Clearly, matrices $T_{k}$ satisfy the recursive relationships similar to
\eqref{1}:
\begin{equation}
T_{0}=\{-\lambda\},\qquad T_{k+1}=\left(
\begin{array}
[c]{c|c}%
T_{k} & J_{k}\\\hline
J_{k}^{\prime} & T_{k}%
\end{array}
\right)  . \label{1'}%
\end{equation}
Denote by $Q_{k}$ the submatrix of $T_{k}$ formed by deleting the \emph{upper}
row and the \emph{right} column, and set
\[
\phi_{k}=\mathrm{det}\,Q_{k}.
\]
Denote by $P_{k}$ the submatrix of $T_{k}$ formed by deleting the \emph{lower}
row and the \emph{right} column, and set
\[
\psi_{k}=\mathrm{det}\,P_{k}.
\]
Finally, for any square matrix $B$ of order $n$, denote by $B^{\prime}$ the
matrix obtained by rotating $B$ by $180^{\circ}$. That is, the elements of
$B^{\prime}$ and $B$ are related by
\[
b_{i,j}^{\prime}=b_{n+1-i,n+1-j},\qquad i,j=1,\ldots,n.
\]

\subsubsection{Auxiliary lemmas}

In this section, we prove a few auxiliary statements.

\begin{lemma}
\label{l0} Each matrix $T_{k}$ satisfies $T_{k}=T_{k}^{\prime}$.
\end{lemma}

\textsf{Proof.} The statement follows from recursive relations \eqref{1'} by
induction in $k$. \hfill$\Box$

\begin{lemma}
\label{l1} The following recursive relationship holds:
\begin{equation}
\label{l1'}\chi_{k+1}=(\chi_{k})^{2} - (\phi_{k})^{2}.
\end{equation}

\end{lemma}

\textsf{Proof.} Considering the expression
\[
\mathrm{det}\, T_{k+1}=\sum_{(i_{1}i_{2}\ldots i_{n})\in S_{n}} \mathrm{sgn}%
\,(i_{1}i_{2}\ldots i_{n})\,t_{1,i_{1}}\cdot t_{2,i_{2}}\cdots\, t_{n,i_{n}}
\]
for the determinant $\chi_{k+1}=\mathrm{det}\, T_{k+1}$, where $n=2^{k+1}$,
and using \eqref{1'}, one can see that each term in the product $t_{1,i_{1}}\cdot
t_{2,i_{2}}\cdots\, t_{n,i_{n}}$ either contains elements from the blocks
$T_{k}$ of the matrix $T_{k+1}$ only or \emph{both} elements $1$ from the
blocks $J_{k}=\mathrm{diag}\,\{1,0,\ldots,0\}$ and $J_{k}^{\prime
}=\mathrm{diag}\,\{0,\ldots,0,1\}$. This implies that
\begin{equation}
\label{3}\mathrm{det}\, T_{k+1}= (\mathrm{det}\, T_{k})^{2} -\mathrm{det}\,
Q_{k}\, \mathrm{det}\, S_{k},
\end{equation}
where the submatrix $S_{k}$ of $T_{k}$ is formed by deleting the \emph{lower}
row and the \emph{left} column. Lemma \ref{l0} implies that $S_{k}%
=Q_{k}^{\prime}$ and since $\mathrm{det}\,B^{\prime}=\mathrm{det}\,B$ for any
square matrix $B$, equation \eqref{3} yields $\mathrm{det}\, T_{k+1}=
(\mathrm{det}\, T_{k})^{2} -(\mathrm{det}\, Q_{k})^{2}$, which is equivalent
to \eqref{l1'}. \hfill$\Box$

\begin{lemma}
\label{l2} The following relation holds:
\begin{equation}
\label{4}\phi_{k+1}=-\phi_{k} \psi_{k}.
\end{equation}

\end{lemma}

\textsf{Proof.} Since the matrix $Q_{k+1}$ is obtained from $T_{k+1}$ by
deleting the upper row and the right column, formula \eqref{1'} implies that
$Q_{k+1}$ is an upper block triangular matrix of the form
\[
Q_{k+1}= \left(
\begin{array}
[c]{c|c}%
Q_{k} & B\\\hline
O & C
\end{array}
\right)  ,
\]
where $O$ is a zero matrix and the square matrix $C$ is defined by
\[
C= \left(
\begin{array}
[c]{c|c}%
0 & P_{k}\\\hline
1 & \alpha
\end{array}
\right)  ,
\]
where $0$ is a zero column, and $\alpha$ is a row. Hence, $\mathrm{det}%
\,Q_{k+1}=\mathrm{det}\,Q_{k}\, \mathrm{det}\,C$. Furthermore, using the
Laplace expansion along the first column, $\mathrm{det}\,C= - \mathrm{det}%
\,P_{k}$, hence $\mathrm{det}\,Q_{k+1}=-\mathrm{det}\,Q_{k}\, \mathrm{det}%
\,P_{k}$, which is equivalent to \eqref{4}. \hfill$\Box$

\begin{lemma}
\label{l3} The following relation holds:
\begin{equation}
\label{5}\psi_{k+1}=\chi_{k} \psi_{k}.
\end{equation}

\end{lemma}

\textsf{Proof.} Since the matrix $P_{k+1}$ is obtained from $T_{k+1}$ by
deleting the lower row and the right column, formula \eqref{1'} implies that
$P_{k+1}$ is an upper block triangular matrix of the form
\[
P_{k+1}= \left(
\begin{array}
[c]{c|c}%
T_{k} & B\\\hline
O & P_{k}%
\end{array}
\right)  ,
\]
hence $\mathrm{det}\,P_{k+1}=\mathrm{det}\,T_{k}\, \mathrm{det}\,P_{k}$, which
is equivalent to \eqref{5}. \hfill$\Box$

\begin{lemma}
\label{l5} Chebyshev polynomials satisfy the relation
\[
(U_{k+1})^{2} -U_{k+2}U_{k}=1.
\]

\end{lemma}

\textsf{Proof.} Applying the recursive relation \eqref{urecursive} gives
\[%
\begin{array}
[c]{rcl}%
(U_{k+1})^{2} - U_{k+2}U_{k} & = & (U_{k+1})^{2} - (2\lambda\,U_{k+1}%
-U_{k})U_{k}\\
& = & (2\lambda\,U_{k}-U_{k-1})^{2} -2\lambda(2\lambda\,U_{k}-U_{k-1})U_{k}
+(U_{k})^{2}\\
& = & (U_{k})^{2} -U_{k+1}U_{k-1}.
\end{array}
\]
Hence, the statement follows from $(U_{1})^{2}-U_{2}U_{0}=1$. \hfill$\Box$

\subsubsection{Proof of the theorem}

Since $\psi_{1}=\chi_{0}=-\lambda$, equation \eqref{5} implies that
\[
\psi_{k}=\prod_{i=0}^{k-1} \chi_{i},\qquad k\ge1.
\]
Combining this with \eqref{4} and taking into account that $\phi_{1}=1$, one
obtains
\[
\phi_{k}=(-1)^{k-1}\prod_{i=0}^{k-2} (\chi_{i})^{k-1-i},\qquad k\ge2.
\]
Therefore, formula \eqref{l1'} implies the recursive relationship
\begin{equation}
\label{2}\chi_{k+1}=(\chi_{k})^{2} - \prod_{i=0}^{k-2} (\chi_{i})^{2(k-1-i)},
\quad k\ge2.
\end{equation}
Note that by direct calculation
\begin{equation}
\label{2'}\chi_{0}=-\lambda;\quad\chi_{1}=\lambda^{2}-1;\quad\chi_{2}%
=\lambda^{4}-2\lambda^{2}.
\end{equation}

Since formulas \eqref{2}, \eqref{2'} uniquely define the sequence of
characteristic polynomials $\chi_{k}$, it remains to show that expressions
\eqref{pol} satisfy these formulas.
Indeed, for $N=1,2,3$, formula \eqref{pol} gives the expressions
\[
\chi_{1}(\lambda)=U_{2},\qquad\chi_{2}(\lambda)=U_{1} U_{3},\qquad\chi
_{3}(\lambda)= U_{1}^{2} U_{2} U_{4},
\]
which are compatible with \eqref{2}, \eqref{2'}; here and henceforth, we omit
the argument of Chebyshev polynomials for brevity, i.e.~$U_{i}$ stands for
$U_{i}(-\lambda/2)$.

For larger $N$, rewriting formula \eqref{2} equivalently as
\[
(\chi_{k})^{2}-\chi_{k+1}=\chi_{0}^{2(k-1)}\prod_{j=1}^{k-2}(\chi
_{j})^{2(k-1-j)}%
\]
and substituting expressions $\chi_{0}=U_{1}$ and \eqref{pol} into this
equation, we obtain
\[%
\begin{array}
[c]{rcl}%
\displaystyle\left(  U_{k+1}\prod_{i=0}^{k-1}\bigl(U_{i}\bigr)^{2^{k-i-1}%
}\right)  ^{2} & - & \displaystyle U_{k+2}\prod_{i=0}^{k}\bigl(U_{i}%
\bigr)^{2^{k-i}}\\
& = & \displaystyle(U_{1})^{2(k-1)}\prod_{j=1}^{k-2}\left(  U_{j+1}\prod
_{i=0}^{j-1}\bigl(U_{i}\bigr)^{2^{j-i-1}}\right)  ^{2(k-1-j)},
\end{array}
\]
where $k\geq3$. Upon rearrangement of the left hand side, this can be written as
\[
\bigl((U_{k+1})^{2}-U_{k+2}U_{k}\bigr)\prod_{i=0}^{k-1}\bigl(U_{i}%
\bigr)^{2^{k-i}}=(U_{1})^{2(k-1)}\prod_{j=1}^{k-2}\left(  U_{j+1}\prod
_{i=0}^{j-1}\bigl(U_{i}\bigr)^{2^{j-i-1}}\right)  ^{2(k-1-j)},
\]
which, by Lemma \ref{l5}, is equivalent to
\begin{equation}
\prod_{i=0}^{k-1}\bigl(U_{i}\bigr)^{2^{k-i}}=(U_{1})^{2(k-1)}\prod_{j=1}%
^{k-2}\left(  U_{j+1}\prod_{i=0}^{j-1}\bigl(U_{i}\bigr)^{2^{j-i-1}}\right)
^{2(k-1-j)}. \label{2''}%
\end{equation}
The right hand side of this equation is the product of powers $(U_{i})^{m_{i}%
}$ of Chebyshev polynomials with $i=1,\ldots,k-1$ (note that $U_{0}=1$).
Counting the number of times each factor $U_{i}$ enters the right hand side of
equation \eqref{2''}, one obtains
\[
m_{i}=2(k-i)+\sum_{j=1}^{k-i-2}2^{k-i-j-1}j,\quad i=1,\ldots,k-3;\qquad
m_{k-2}=4;\qquad m_{k-1}=2.
\]
It is easy to see by induction that
\[
2n+\sum_{j=1}^{n-2}2^{n-j-1}j=2^{n}%
\]
for all $n\geq3$; hence, $m_{i}=2^{k-i}$ for all $i=1,\ldots,k-1$, and
therefore, \eqref{2''} is an identity. This establishes that equations
\eqref{pol} satisfy recursive relations \eqref{2}, \eqref{2'} and completes
the proof of the theorem.

\section{Eigenvalue distribution of the adjacency matrix}

\label{section_eigenvalues}In this section, we consider the distribution
function (empirical spectral distribution, density of states)%
\begin{equation}
F_{N}(x)=\frac{\#\left\{  \lambda\in\mathrm{sp\,}\left(  \frac{1}{2}%
A_{N}\right)  :\lambda<x\right\}  }{2^{N}}, \label{dist}%
\end{equation}
of the eigenvalues of the matrix $\frac{1}{2}A_{N}$ and its $N\rightarrow
\infty$ limit (limiting spectral distribution).
By definition, $F_{N}$ is an increasing piecewise constant left-continuous
function with a finite number of jumps. Since the spectrum of the matrix
$\frac{1}{2}{A_{N}}$ belongs to the interval $(-1,1)$, it also follows that
$F_{N}(x)=0$ for $x\leq-1$ and $F_{N}(x)=1$ for $x\geq1$.

The following Devil's staircase function
\cite{bay,bom,davison} has received substantial attention in the literature:
\begin{equation}
f(x)=\sum_{k=1}^{\infty}\frac{\lfloor kx\rfloor}{2^{k}},\qquad0\leq x<1,
\label{devil}%
\end{equation}
where $\lfloor\cdot\rfloor$ denotes the \emph{floor} (integer value) function. We
consider the extension of $f$ to the whole axis according to the formulas
\begin{equation}
f(x)=0\quad\mathrm{for}\quad x<0;\qquad f(x)=1\quad\mathrm{for}\quad x\geq1.
\label{devil1}%
\end{equation}
This extension is an increasing right continuous function. At every rational
point $x_{i}=r/q\in(0,1)$, where $r,q>0$ are coprime integers, the function
$f$ has a jump
\[
\Delta f(x_{i})=f(x_{i}+0)-f(x_{i}-0)=\frac{1}{2^{q}-1},\qquad x_{i}=\frac
{r}{q}\in(0,1),
\]
and $f$ is continuous at every irrational point. As a matter of fact,
equation \eqref{devil} is equivalent to
\begin{equation}
f\left(  x\right)  =%
\begin{cases}
\displaystyle\sum_{p=1}^{\infty}\frac{1}{2^{\left\lfloor \frac{p}%
{x}\right\rfloor }} & \text{for irrational $x$},\\
\displaystyle\sum_{p=1}^{\infty}\frac{1}{2^{\left\lfloor \frac{p}%
{x}\right\rfloor }}+\frac{1}{2^{q}-1} & \text{for rational $x=\frac{r}{q}$}\,.
\end{cases}
\label{devil'}%
\end{equation}
Further, the total sum of all the jumps of $f$ satisfies%
\[
\sum_{i}\Delta f(x_{i})=\sum_{q=2}^{\infty}\frac{\varphi(q)}{2^{q}-1}=1,
\]
where $\varphi$ is Euler's totient function. Hence, this sum equals the total
variation $f(1)-f(0)=1$ of $f$, and therefore $f$ is a jump function
\cite{ambrosio2000functions}, i.e.
\[
f(x)=\sum_{x_{i}<x}\Delta f(x_{i}),\qquad x\in\mathbb{R},
\]
with the sum over all the rational points $x_{i}\in(0,x)$.

\begin{theorem}
\label{t3} The distribution function of the spectrum of the matrix $A_{N}$
satisfies the limit relationship
\begin{equation}
\label{limit}\lim_{N\to\infty} F_{N}(x)=1-f\left(  \frac1\pi\arccos x \right)
,\qquad-1\le x\le1,
\end{equation}
where $f$ is defined by \eqref{devil}.
\end{theorem}

This theorem is proved in the next section. From formulas \eqref{pol},
\eqref{pol1}, it follows that the distribution function of the spectrum of the
matrix $\frac{1}{2}A_{N}^{\prime}$ converges to the same limit as
$N\rightarrow\infty$. It should be noted that the algebraic and geometric
multiplicities of eigenvalues are different.

\subsection{Proof of Theorem \ref{t3}}

Let us order the zeros of the
polynomial
\begin{equation}
\frac{\chi_{n}(\lambda)}{U_{n+1}(\lambda/2)}=\prod_{q=1}^{n}(U_{q-1}\left(
\lambda/2\right)  )^{2^{n-q}} \label{div}%
\end{equation}
(cf.~\eqref{pol}) in a $(n-1)\times n$
matrix%
\[
D=\left(
\begin{array}
[c]{cccccc}%
0 & 2^{n-2} & 2^{n-3} & 2^{n-4} & \cdots & 2^{0}\\
0 & 0 & 2^{n-3} & 2^{n-4} & \cdots & 2^{0}\\
0 & 0 & 0 & 2^{n-4} & \cdots & 2^{0}\\
\vdots & \vdots & \vdots & \vdots & \ddots & \vdots\\
0 & 0 & 0 & 0 & \cdots & 2^{0}%
\end{array}
\right)  ,
\]
where at position $(p,q)$ with $1\leq p\leq n-1$, $1\leq q\leq n$, we put down
the entry $D_{p,q}=2^{n-q}$ which corresponds to the multiplicity of the
factor $U_{q-1}$ in \eqref{div}. For every $x$ and $n$ we define the set of
pairs
\[
A_{n}\left(  x\right)  =\left\{  \left(  p,q\right)  :\ {p}/{q}\leq
x\wedge1<q\leq n\right\}
\]
and the function
\[
f_{n}\left(  x\right)  =\sum_{\left(  p,q\right)  \in A_{n}\left(  x\right)
}D_{p,q}=\sum_{\left(  p,q\right)  \in A_{n}\left(  x\right)  }\frac{1}{2^{q}%
}.
\]
Equation \eqref{zeros} implies that this function and the distribution
function \eqref{dist} are related by
\begin{equation}
F_{n}(x)=1-f_{n}\left(  \frac{1}{\pi}\arccos x\right)  . \label{rel}%
\end{equation}
Now note that
\[
A_{n}\left(  x\right)  =\bigcup_{p=1}^{n-1}B_{n,p}\left(  x\right)
\]
with
\[
B_{n,p}\left(  x\right)  =\left\{  \left(  p,q\right)  :\ {p}/{x}\leq q\leq
n\right\}
\]
and therefore,
\[
\sum_{\left(  p,q\right)  \in B_{n,p}\left(  x\right)  }\frac{1}{2^{q}}%
=\frac{1}{2^{\left\lceil \frac{p}{x}\right\rceil }}+\frac{1}{2^{\left\lceil
\frac{p}{x}\right\rceil +1}}+\ldots+\frac{1}{2^{n}}
=2^{1-\left\lceil \frac{p}{x}\right\rceil }-2^{-n}.
\]
Summing this over $p$ gives
\[
f_{n}\left(  x\right)  =\sum_{p=1}^{n-1}\sum_{\left(  p,q\right)  \in
B_{n,p}\left(  x\right)  }\frac{1}{2^{q}}=\sum_{p=1}^{n-1}\left[
2^{1-\left\lceil \frac{p}{x}\right\rceil }-2^{-n}\right]  =\left[  \sum
_{p=1}^{n-1}\frac{1}{2^{\left\lceil \frac{p}{x}\right\rceil -1}}\right]
-\left(  n-1\right)  2^{-n}.
\]
In the limit $n\rightarrow\infty$ the last term vanishes and we obtain
\begin{equation}
\lim_{n\rightarrow\infty}f_{n}\left(  x\right)  =\sum_{p=1}^{\infty}\frac
{1}{2^{\left\lceil \frac{p}{x}\right\rceil -1}}, \label{limi}%
\end{equation}
where $\lceil\cdot\rceil$ is the \emph{ceiling} function. Finally, note that
$\left\lceil \frac{p}{x}\right\rceil -1=\left\lfloor \frac{p}{x}\right\rfloor
$ for all $p$ if $x$ is irrational, while if $x=r/q$ is a rational number then
$\left\lceil \frac{p}{x}\right\rceil -1=\left\lceil \frac{pq}{r}\right\rceil
-1=\left\lfloor \frac{p}{x}\right\rfloor -1$ whenever $p=mr$ is a multiple of
$r$. We thus see that
the limit \eqref{limi} coincides with the function \eqref{devil'} and hence
\eqref{rel} implies \eqref{limit}, which completes the proof.

\section{Eigenvectors}

\label{section_eigenvectors}Eigenvectors and eigenspaces of graphs are also
important \cite{cvetkovic1997eigenspaces}, \cite{van2010graph}. For example
they are used to count walks in a graph and to relate the symmetry of a graph
to its spectrum. The principal eigenvector of the adjacency matrix provides
vertex centrality information, while the second eigenvector can be used to
partition the graph into clusters. The eigenvectors are related to the graph's
automorphic structure.

Let us now study the eigenvectors of $\frac{1}{2}A_{N}$. In order to do this,
let us first make the notation for vertices more efficient. In the following,
let us identify a vertex with the corresponding string of $0$s and $1$s, for
example,
\[
\mathbf{x}=\left(  1,1,0,0,1\right)  =\left(  11001\right)  .
\]
Strings are concatenated by simply writing one after the other. We will use
the notation $1^{j}$ to denote a string with $j$ ones and $0^{j}$ for a string
with $j$ zeros. For example, $\left(  11001\right)  =\left(  1^{2}%
0^{2}1\right)  $. For consistency, it is convenient to agree that $0^{0}$ and
$1^{0}$ is an empty string. Also for convenience, let us introduce the
symmetry operation $S$, which replaces every $0$ in a string with a $1$ and
vice versa, for example, $S\left(  11001\right)  =\left(  00110\right)  $.
Obviously, $S$ is an involution.

An arbitrary binary string of length $j$ will be denoted by
$q^{j}$, and the set of all strings of length $j$ will be denoted by $W_{j}$. In
particular, the matrix $\frac{1}{2}A_{N}$ acts in the $2^{N}$-dimensional
vector space $V$ spanned by the vertices $(q^{N})\in W_{N}$.

Consider an expansion of an eigenvector $v\in V$ of $\frac{1}{2}A_{N}$ along
the basis $W_{N}$:
\begin{equation}
v=\sum_{(q^{N})\in W_{N}}c_{(q^{N})}\left(  q^{N}\right)  , \qquad\frac{1}%
{2}A_{N}v=\lambda v . \label{eq:ev}%
\end{equation}
According to Theorem \ref{t1}, $\lambda$ is a root of a Chebyshev polynomial
$U_{\ell}$ with $\ell=1,\ldots,N-1$ or $\ell=N+1$. We would like to determine
the $2^{N}$ components $c_{(q^{N})}$ of an eigenvector $v$ (up to a scaling factor).

\begin{theorem}
\label{t4} Suppose that $\lambda$ is a root of a Chebyshev polynomial
$U_{\ell+1}$ with $\ell\le N-2$ and is not a root of the Chebyshev polynomials
$U_{1},\ldots,U_{\ell}$. Then, for every $(q^{N-\ell-2})\in W_{N-\ell-2}$, the
following relations define an eigenvector of the matrix $\frac12 A_{N}$:

\begin{itemize}
\item[\textrm{(i)}] $c_{(q^{N-\ell-2}00p^{\ell})}
=0$\quad\textit{for\ all}\quad$(p^{\ell})\in W_{\ell}$;

\item[\textrm{(ii)}] $c_{(q^{N-\ell-2}100^{\ell})}=1$;

\item[\textrm{(iii)}] For every $p^{\ell}=0^{j_{k+1}}1^{j_{k}}0^{j_{k-1}%
}1^{j_{k-2}}\cdots0^{j_{2}}1^{j_{1}}$ with $j_{1}+\cdots+j_{k+1}=\ell$ and
$j_{1}, j_{k+1}\ge0$,\ $j_{2},\ldots,j_{k}\ge1$,
\begin{equation}
\label{cmainmax}c_{(q^{N-\ell-2}10p^{\ell})}\prod_{i=1}^{k} U_{s_{i}}%
(\lambda)=1,
\end{equation}
where $s_{i}=j_{1}+\cdots+j_{i}$;

\item[\textrm{(iv)}] $c_{(q^{N-\ell-2}11p^{\ell})}=0$ and
\begin{equation}
\label{cmainmax'}c_{(q^{N-\ell-2}01p^{\ell})}=-U_{\ell}(\lambda)c_{(q^{N-\ell
-2}10 S(p^{\ell}))}%
\end{equation}
{for all} $(p^{\ell})\in W_{\ell}$.
\end{itemize}

Similarly, if $\lambda$ is a root of the Chebyshev polynomial $U_{N+1}$ and is
not a root of the Chebyshev polynomials $U_{1},\ldots,U_{N-1}$, then the
following relations define an eigenvector of the matrix $\frac12 A_{N}$ with
the eigenvalue $\lambda$:

\begin{itemize}
\item[\textrm{(j)}] $c_{(0^{N})}=1$;

\item[\textrm{(jj)}] For every $q^{N}=0^{j_{k+1}}1^{j_{k}}0^{j_{k-1}%
}1^{j_{k-2}}\cdots0^{j_{2}}1^{j_{1}}$ with $j_{1}+\cdots+j_{k+1}=N$ and
$j_{1}\ge0$,\ $j_{2},\ldots,j_{k},j_{k+1}\ge1$,
\begin{equation}
\label{cmainmax''}c_{(q^{N})}\prod_{i=1}^{k} U_{s_{i}}(\lambda)=1,
\end{equation}
where $s_{i}=j_{1}+\cdots+j_{i}$;

\item[\textrm{(jjj)}] $c_{S(q^{N})}=U_{N}(\lambda)c_{(q^{N})}$\quad for
all\quad$(q^{N})\in W_{N}$.
\end{itemize}
\end{theorem}

Note that the identity $U_{k}^{2}-1=U_{k-1}U_{k+1}$ implies $U_{\ell}%
^{2}(\lambda)=1$ under the conditions of Theorem \ref{t4} because $U_{\ell
+1}(\lambda)=0$. Since $(q^{N-\ell-2})\in W_{N-\ell-2}$ is arbitrary, the
geometric multiplicity of such an eigenvalue is $2^{N-\ell-2}$.

As an example, if $\lambda=1/\sqrt{2}$, then $U_{3}(\lambda)=0$ and
$U_{1}(\lambda)=\sqrt{2}, U_{2}(\lambda)=1$. According to Theorem \ref{t4},
$2^{N-4}$ linearly independent eigenvectors corresponding to the eigenvalue
$\lambda=1/\sqrt{2}$ can be labeled by the strings $(q^{N-4})\in W_{N-4}$, and
the components of an eigenvector are defined by
\[%
\begin{array}
[c]{c}%
c_{(q^{N-4}0000)}=c_{(q^{N-4}0001)}=c_{(q^{N-4}0010)}=c_{(q^{N-4}0011)}=0;\\
c_{(q^{N-4}1111)}=c_{(q^{N-4}1100)}=c_{(q^{N-4}1101)}=c_{(q^{N-4}1100)}=0;\\
~\\
c_{(q^{N-4}1000)}=1, \ c_{(q^{N-4}1001)}=\frac{1}{u_{1}}, \ c_{(q^{N-4}%
1011)}=\frac1{u_{2}}, \ c_{(q^{N-4}1010)}=\frac1{u_{1}u_{2}};\\
~\\
c_{(q^{N-4}1000)}=-u_{2}, \ c_{(q^{N-4}1001)}=-\frac{u_{2}}{u_{1}},
\ c_{(q^{N-4}1011)}=-1, \ c_{(q^{N-4}1010)}=-\frac1{u_{1}},
\end{array}
\]
where $u_{1}=U_{1}(\lambda)=\sqrt{2}$, $u_{2}=U_{2}(\lambda)=1$.

As an illustration of formulas \eqref{cmainmax}, \eqref{cmainmax'} for a
larger $m$, if, for example, $m=10$ and $(p^{m})=0100100110$, then
\[
c_{(q^{N-m-2}10p^{m})}=\frac1{u_{9}u_{8}u_{6}u_{5}u_{3}u_{1}},\qquad
c_{(q^{N-m-2}01p^{m})}=-\frac{u_{10}}{u_{9}u_{8}u_{6}u_{5}u_{3}u_{1}},
\]
where $u_{k}=U_{k}(\lambda)$ and, in particular, $u_{10}^{2}=1$.

Now, let us consider the matrix $\frac12 A_{N}^{\prime}$.

\begin{theorem}
\label{t5} The matrices $\frac12 A_{N}$ and $\frac12 A_{N}^{\prime}$ have the
same eigenvectors (defined by \textrm{(i) -- (iv)}) for each eigenvalue
$\lambda$, which is a root of a Chebyshev polynomial $U_{m}$ with $m\le N-1$.

If $\lambda$ is a root of the Chebyshev polynomial $U_{N}$ and is not a root
of the Chebyshev polynomials $U_{1},\ldots,U_{N-1}$, then the eigenvector of
the matrix $\frac12 A_{N}^{\prime}$ corresponding to the eigenvalue $\lambda$
is defined by relations \textrm{(j), (jj)} of Theorem \ref{t4} and the
equality
\[
c_{S(q^{N})}=-U_{N-1}(\lambda)c_{(q^{N})},\qquad(q^{N})\in W_{N}.
\]

Finally, the components of the eigenvector corresponding to the eigenvalue $1$
are defined by the relations
\[
c_{(q^{N})}\prod_{i=1}^{k} \left(  1+\sum_{m=1}^{i} j_{i}\right)  =1,\qquad
c_{S(q^{N})}=c_{(q^{N})},
\]
for every $q^{N}=0^{j_{k+1}}1^{j_{k}}0^{j_{k-1}}1^{j_{k-2}}\cdots0^{j_{2}%
}1^{j_{1}}$ with $j_{1}+\cdots+j_{k+1}=N$ and $j_{1}\ge0$,\ $j_{2}%
,\ldots,j_{k},j_{k+1}\ge1$.
\end{theorem}

Theorem \ref{t4} is proved in the next section. The proof of Theorem \ref{t5}
follows the same line and is omitted.

\subsection{Proof of Theorem \ref{t4}}

We again use the notation $u_{j}=U_{j}(\lambda)$. The proof is based on the
following lemma.

\begin{lemma}
\label{l6} For every $m\le N-2$, every $q^{N-m-2}\in W_{N-m-2}$ and every
$1\le j\le m$, the components of an eigenvector of $\frac12 A_{N}$ with an
eigenvalue $\lambda$ satisfy
\begin{equation}
\label{aa1}c_{(q^{N-m-2}01^{m+1-j}0^{j})}+u_{j-1}\sum_{i=j}^{m} c_{(q^{N-m-2}%
01^{m-i}01^{i})}=u_{j} c_{(q^{N-m-2}01^{m+1})},
\end{equation}
\begin{equation}
\label{aa2}c_{(q^{N-m-2}0^{m+2})}=u_{m+1} c_{(q^{N-m-2}01^{m+1})},
\end{equation}
\begin{equation}
\label{aa3}c_{(q^{N-m-2}10^{m+1-j}1^{j})}+u_{j-1}\sum_{i=j}^{m} c_{(q^{N-m-2}%
10^{m-i}10^{i})}=u_{j} c_{(q^{N-m-2}10^{m+1})},
\end{equation}
\begin{equation}
\label{aa4}c_{(q^{N-m-2}1^{m+2})}=u_{m+1} c_{(q^{N-m-2}10^{m+1})}%
\end{equation}
provided that $u_{j}\ne0$ for $j=1,\ldots,m-1$. Further, if $u_{j}\ne0$ for
$j=1,\ldots, N-2$, then
\begin{equation}
\label{aa5}c_{(1^{N-j}0^{j})}+u_{j-1}\sum_{i=j}^{N-1} c_{(1^{N-1-i}01^{i}%
)}=u_{j} c_{(1^{N})},
\end{equation}
\begin{equation}
\label{aa5'}c_{(0^{N-j}1^{j})}+u_{j-1}\sum_{i=j}^{N-1} c_{(1^{N-1-i}10^{i}%
)}=u_{j} c_{(0^{N})},%
\end{equation}
for all $1\le j\le N-1$, and
\begin{equation}
\label{aa5''}c_{(0^{N})}=u_{N} c_{(1^{N})},\qquad c_{(1^{N})}=u_{N}
c_{(0^{N})}.
\end{equation}

\end{lemma}

\textsf{Proof.} By the definition of $\frac12A_{N}$, for every eigenvector of this
matrix with an eigenvalue $\lambda$, we have
\[
c_{(q^{N-2}0^{2})}=u_{1} c_{(q^{N-2}01)},\qquad c_{(q^{N-2}0^{2})}=u_{1}
c_{(q^{N-2}01)},
\]
\[
\sum_{j=0}^{m} c_{(q^{N-m-2}01^{m-j}01^{j})}=u_{1} c_{(q^{N-m-2}01^{m+1})},
\]
\[
\sum_{j=0}^{m} c_{(q^{N-m-2}10^{m-j}10^{j})}=u_{1} c_{(q^{N-m-2}10^{m+1})}
\]
for all $q^{N-2}\in W_{N-2}$, $q^{N-m-2}\in W_{{N-m-2}}$ (note that
$u_{1}=2\lambda$). These relations coincide with \eqref{aa2}, \eqref{aa4} for
$m=0$ and with \eqref{aa1}, \eqref{aa3} for any $1\le m\le N-2$, $j=1$,
respectively, and can be used as the basis for the induction in $m, j$. Due to
$S$-symmetry, every formula obtained below remains valid if we replace all
$0^\prime$s with $1^\prime$s and vice versa.

For the induction step, let $1\le\hat m\le N-2$ and assume that \eqref{aa2},
\eqref{aa4} hold for $m\le\hat m-1$ and \eqref{aa1}, \eqref{aa3} hold for
$m=\hat m$ and $j\le\hat m-1$.

Multiplying equation \eqref{aa1} with $m=\hat m$ by $u_{j}$ and using
\eqref{aa2}, \eqref{aa4} with $m=j-1$, we obtain
\[%
\begin{array}
[c]{rcl}%
c_{(q^{N-\hat m-2}01^{\hat m+1})}+u_{j-1}c_{(q^{N-\hat m-2}01^{\hat
m-j}0^{j+1})} & + & \displaystyle u_{j-1} u_{j}\sum_{i=j+1}^{\hat m}
c_{(q^{N-\hat m-2}01^{\hat m-i}01^{i})}\\
~ &  & \\
& = & u_{j}^{2} c_{(q^{N-\hat m-2}01^{\hat m+1})}.
\end{array}
\]
Since $u_{j}^{2}-1=u_{j-1}u_{j+1}$, this is equivalent to
\[
c_{(q^{N-\hat m-2}01^{\hat m-j}0^{j+1})} + u_{j}\sum_{i=j+1}^{\hat m}
c_{(q^{N-\hat m-2}01^{\hat m-i}01^{i})}\newline~\newline=u_{j+1} c_{(q^{N-\hat
m-2}01^{\hat m+1})}
\]
provided that $u_{j-1}\ne0$, which is equivalent to \eqref{aa1} with $m=\hat
m$ and $j$ replaced with $j+1$. By induction, this proves equation \eqref{aa1}
(and, similarly, \eqref{aa3}) for all $j\le\hat m$.

Further, setting $j=m=\hat m$ in \eqref{aa1} gives
\[
c_{(q^{N-\hat m-2}010^{\hat m})}+u_{\hat m-1} c_{(q^{N-\hat m-2}001^{\hat m}%
)}=u_{\hat m} c_{(q^{N-\hat m-2}01^{\hat m+1})}.
\]
Multiplying with $u_{\hat m}$ and using \eqref{aa2}, \eqref{aa4} with $m=\hat
m-1$, we obtain
\[
c_{(q^{N-\hat m-2}01^{\hat m+1})}+u_{\hat m-1} c_{(q^{N-\hat m-2}0^{\hat
m+2})}=u_{\hat m}^{2} c_{(q^{N-\hat m-2}01^{\hat m+1})},
\]
which, due to $u_{\hat m}^{2}-1=u_{\hat m-1}u_{\hat m+1}$, is equivalent to
\eqref{aa1} with $m=\hat m$, provided that $u_{\hat m-1}\ne0$. The proof of
\eqref{aa4} with $m=\hat m$ is similar. This completes the induction step and
the proof of \eqref{aa1} -- \eqref{aa4}. Formulas \eqref{aa5} -- \eqref{aa5''}
can be obtained in the exact same manner. \hfill$\Box$

\medskip The conclusions of the theorem follow easily from Lemma
\ref{l6}. In particular, if $u_{\ell+1}=0$ and $u_{k}\ne0$ for $k\le\ell$,
then by successively applying formulas \eqref{aa2}, \eqref{aa4} with
$m+1=s_{k},s_{k-1},\ldots,s_{1}$, we obtain
\begin{equation}
\label{pro}c_{(q^{N-\ell-2}10p^{\ell})}\prod_{i=1}^{k} u_{s_{i}}%
=c_{(q^{N-\ell-2}10^{\ell+1})},\quad c_{(q^{N-\ell-2}00p^{\ell})}\prod
_{i=1}^{k} u_{s_{i}}=c_{(q^{N-\ell-2}0^{\ell+2})},
\end{equation}
\begin{equation}
\label{pro'}c_{(q^{N-\ell-2}01p^{\ell})}\prod_{i=1}^{k} u_{s_{i}%
}=c_{(q^{N-\ell-2}01^{\ell+1})},\quad c_{(q^{N-\ell-2}11p^{\ell})}\prod
_{i=1}^{k} u_{s_{i}}=c_{(q^{N-\ell-2}1^{\ell+2})}.
\end{equation}
Further, setting $m=\ell$ in \eqref{aa2}, \eqref{aa4} and taking into account
that $u_{\ell+1}=0$ gives
\begin{equation}
\label{prox}c_{(q^{N-\ell-2}0^{\ell+2})}= c_{(q^{N-\ell-2}1^{\ell+2})}=0.
\end{equation}
On the other hand, formulas \eqref{aa1}, \eqref{aa3} with $j=m=\ell+1$ result
in the equations
\[
c_{(q^{N-\ell-3}010^{\ell+1})}+u_{\ell} c_{(q^{N-\ell-3}001^{\ell+1}%
)}=0,\ \ \ c_{(q^{N-\ell-3}101^{\ell+1})}+u_{\ell} c_{(q^{N-\ell-3}%
110^{\ell+1})}=0,
\]
which, due to $u_{\ell}^{2}=1$, can be combined into one equation
\begin{equation}
\label{prox'}c_{(q^{N-\ell-2}10^{\ell+1})}=-u_{\ell} c_{(q^{N-\ell-3}%
01^{\ell+1})}.
\end{equation}
Combining relations \eqref{pro} -- \eqref{prox'} with the normalization
condition $c_{(q^{N-\ell-2}100^{\ell})}=1$, one obtains statements (i) -- (iv)
of the theorem. The proof of statements (j) -- (jjj) follows from relations
\eqref{aa5} -- \eqref{aa5''} of Lemma \ref{l6} in a similar fashion.

\section{Conclusions}

\label{section_conclusions}We considered the graph describing the transitions
between the discrete states of the Preisach input-state-output hysteresis
model. The graph has a self-similar (block-hierarchical) non-symmetric adjacency matrix. Its
eigenvalues, their multiplicities (eigenvalue distribution), and their
corresponding eigenvectors were explicitly calculated. These can be used to
glean information about the underlying walk, for example, in describing the
invariant distribution. These results are expected to prove useful in
identifying parameters of the Preisach model (the Preisach measure). Our
approach can also be fruitful in studying other systems: in particular, complex
networks where the connection structure is self-similar.

\subsection*{Acknowledgments}

The authors thank M. Arnold for a stimulating discussion of the results. D.R.
acknowledges the support of NSF through grant DMS-$1413223$.

\bibliographystyle{abbrv}
\bibliography{hystereticspectrum}

\begin{thebibliography}{10}

\bibitem{amann2012characterization}
A.~Amann, M.~Brokate, S.~McCarthy, D.~Rachinskii, and G.~Temnov.
\newblock Characterization of memory states of the {P}reisach operator with
  stochastic inputs.
\newblock {\em Physica B: Condensed Matter}, 407(9):1404--1411, 2012.

\bibitem{ambrosio2000functions}
L.~Ambrosio, N.~Fusco, and D.~Pallara.
\newblock {\em Functions of bounded variation and free discontinuity problems}.
\newblock Oxford University Press, 2000.

\bibitem{aubry}
S.~Aubry.
\newblock Exact models with a complete {D}evil's staircase.
\newblock {\em J. Phys. C: Solid State Phys.}, 16:2497--2508, 1983.

\bibitem{bay}
D.~H. Bailey and R.~E. Crandall.
\newblock Random generators and normal numbers.
\newblock {\em Exper. Math.}, 11:527--546, 2002.

\bibitem{barriere2009generalized}
L.~Barri{\`e}re, C.~Dalf{\'o}, M.~A. Fiol, and M.~Mitjana.
\newblock The generalized hierarchical product of graphs.
\newblock {\em Discrete Mathematics}, 309(12):3871--3881, 2009.

\bibitem{bax}
R.~J. Baxter.
\newblock {\em Exactly solved models in statistical mechanics}.
\newblock Academic Press, 1982.

\bibitem{bom}
P.~E. B\"ohmer.
\newblock \"uber die transcendenz gewisser dyadischer br\"uche.
\newblock {\em Math. Ann.}, 96:367--377, 1926.

\bibitem{and}
C.~J. Bolech and N.~Andrei.
\newblock Solution of the two-channel anderson impurity model: Implications for
  the heavy fermion {UB}e${}_{13}$.
\newblock {\em Phys. Rev. Lett.}, 88:237206, 2002.

\bibitem{bottcher2005spectral}
A.~B{\"o}ttcher and S.~M. Grudsky.
\newblock {\em Spectral properties of banded {T}oeplitz matrices}.
\newblock SIAM, 2005.

\bibitem{brokate2012hysteresis}
M.~Brokate and J.~Sprekels.
\newblock {\em Hysteresis and phase transitions}, volume 121.
\newblock Springer Science \& Business Media, 2012.

\bibitem{brokate1989properties}
M.~Brokate and A.~Visintin.
\newblock Properties of the {P}reisach model for hysteresis.
\newblock {\em J. Reine Angew. Math.}, 402:1--40, 1989.

\bibitem{chung1997spectral}
F.~R. Chung.
\newblock {\em Spectral graph theory}, volume~92.
\newblock American Mathematical Soc., 1997.

\bibitem{cvetkovic1997eigenspaces}
D.~M. Cvetkovi{\'c}, P.~Rowlinson, and S.~Simic.
\newblock {\em Eigenspaces of graphs}, volume~66.
\newblock Cambridge University Press, 1997.

\bibitem{davison}
J.~L. Davison.
\newblock A series and its associated continued fraction.
\newblock {\em Proc. Amer. Math. Soc.}, 63:29--32, 1977.

\bibitem{dim}
M.~Dimian and P.~Andrei.
\newblock {\em Noise-Driven Phenomena in Hysteretic Systems}.
\newblock Springer, 2013.

\bibitem{duits2008equilibrium}
M.~Duits and A.~B. Kuijlaars.
\newblock An equilibrium problem for the limiting eigenvalue distribution of
  banded {T}oeplitz matrices.
\newblock {\em SIAM J. Matrix Analysis and Applications}, 30(1):173--196, 2008.

\bibitem{ferrand2007analogue}
E.~Ferrand.
\newblock An analogue of the {T}hue-{M}orse sequence.
\newblock {\em The Electronic Journal of Combinatorics}, 14(1):R30, 2007.

\bibitem{fu1997perfect}
X.~Fu, Y.~Liu, P.~Zhou, and W.~Sritrakool.
\newblock Perfect self-similarity of energy spectra and gap-labeling properties
  in one-dimensional {F}ibonacci-class quasilattices.
\newblock {\em Physical Review B}, 55(5):2882, 1997.

\bibitem{gutkin2011spectral}
B.~Gutkin and V.~A. Osipov.
\newblock Spectral problem of block-rectangular hierarchical matrices.
\newblock {\em Journal of Statistical Physics}, 143(1):72--87, 2011.

\bibitem{he2000laplacian}
L.~He, X.~Liu, and G.~Strang.
\newblock Laplacian eigenvalues of growing trees.
\newblock In {\em Conf. on Math. Theory of Networks and Systems}, 2000.

\bibitem{he2003trees}
L.~He, X.~Liu, and G.~Strang.
\newblock Trees with {C}antor eigenvalue distribution.
\newblock {\em Studies in Applied Mathematics}, 110(2):123--138, 2003.

\bibitem{hirschman1967spectra}
I.~Hirschman et~al.
\newblock The spectra of certain {T}oeplitz matrices.
\newblock {\em Illinois Journal of Mathematics}, 11(1):145--159, 1967.

\bibitem{hsu1993fibonacci}
W.-J. Hsu and J.-S. Liu.
\newblock Fibonacci cubes - a class of self-similar graphs.
\newblock In {\em Fibonacci Quart}. Citeseer, 1993.

\bibitem{katsanos2001level}
D.~Katsanos and S.~Evangelou.
\newblock Level-spacing distribution of a fractal matrix.
\newblock {\em Physics Letters A}, 289(4):183--187, 2001.

\bibitem{komjathy2011generating}
J.~Komj{\'a}thy and K.~Simon.
\newblock Generating hierarchial scale-free graphs from fractals.
\newblock {\em Chaos, Solitons \& Fractals}, 44(8):651--666, 2011.

\bibitem{may}
C.~E. Korman and I.~D. Mayergoyz.
\newblock Preisach model driven by stochastic inputs as a model for
  aftereffect.
\newblock {\em IEEE Transactions on Magnetics}, 32(5), 1996.

\bibitem{kostadinov1987fractal}
I.~Kostadinov.
\newblock Fractal {H}amiltonians in condensed matter physics.
\newblock {\em Physica Scripta}, 36(3):516, 1987.

\bibitem{krasnosel2012systems}
M.~A. Krasnosel'skii and A.~V. Pokrovskii.
\newblock {\em Systems with hysteresis}.
\newblock Springer Science \& Business Media, 2012.

\bibitem{krejci}
P.~Krej\v{c}\'\i.
\newblock {\em Hysteresis, convexity and dissipation in hyperbolic equations}.
\newblock Gattotoscho, 1996.

\bibitem{arxiv}
P.~Krej\v{c}\i, J.~P. O'Kane, A.~Pokrovskii, and D.~Rachinskii.
\newblock Properties of solutions to a class of differential models
  incorporating preisach hysteresis operator.
\newblock {\em Physica D: Nonlinear Phenomena}, 241(22):2010--2028, 2012.

\bibitem{leskovec2010kronecker}
J.~Leskovec, D.~Chakrabarti, J.~Kleinberg, C.~Faloutsos, and Z.~Ghahramani.
\newblock {K}ronecker graphs: An approach to modeling networks.
\newblock {\em Journal of Machine Learning Research}, 11(Feb):985--1042, 2010.

\bibitem{mason2002chebyshev}
J.~C. Mason and D.~C. Handscomb.
\newblock {\em Chebyshev polynomials}.
\newblock CRC Press, 2002.

\bibitem{science}
I.~Mayergoyz and G.~Bertotti, editors.
\newblock {\em The science of hysteresis}.
\newblock Academic Press, 2005.

\bibitem{mayergoyz2003mathematical}
I.~D. Mayergoyz.
\newblock {\em Mathematical models of hysteresis and their applications}.
\newblock Academic Press, 2003.

\bibitem{preisach1935magnetische}
F.~Preisach.
\newblock {\"U}ber die magnetische nachwirkung.
\newblock {\em Zeitschrift f{\"u}r Physik}, 94(5-6):277--302, 1935.

\bibitem{siam}
D.~Rachinskii and M.~Ruderman.
\newblock Convergence of direct recursive algorithm for identification of
  preisach hysteresis model with stochastic input.
\newblock {\em SIAM J. Appl. Math.}, 76(4):1270--1295, 2016.

\bibitem{radons2008spectral}
G.~Radons.
\newblock Spectral properties of the preisach hysteresis model with random
  input. i. general results.
\newblock {\em Physical Review E}, 77(6):061133, 2008.

\bibitem{schmidt1960toeplitz}
P.~Schmidt and F.~Spitzer.
\newblock The {T}oeplitz matrices of an arbitrary {L}aurent polynomial.
\newblock {\em Math. Scand}, 8(15-38):11J, 1960.

\bibitem{radons1}
S.~Schubert and G.~Radons.
\newblock Preisach models of hysteresis driven by markovian input processes.
\newblock {\em Physical Review E}, 96(2):022117, 2017.

\bibitem{sethna}
J.~P. Sethna, K.~Dahmen, S.~Kartha, J.~A. Krumhansl, B.~W. Roberts, and J.~D.
  Shore.
\newblock Hysteresis and hierarchies: Dynamics of disorder-driven first-order
  phase transformations.
\newblock {\em Phys. Rev. Lett.}, 70:3347--3350, 1993.

\bibitem{song2005self}
C.~Song, S.~Havlin, and H.~A. Makse.
\newblock Self-similarity of complex networks.
\newblock {\em Nature}, 433(7024):392--395, 2005.

\bibitem{stosic1997residual}
B.~D. Stosic, T.~Stosic, I.~P. Fittipaldi, and J.~Veerman.
\newblock Residual entropy of the square {I}sing antiferromagnet in the maximum
  critical field: the {F}ibonacci matrix.
\newblock {\em Journal of Physics A: Mathematical and General}, 30(10):L331,
  1997.

\bibitem{van2010graph}
P.~Van~Mieghem.
\newblock {\em Graph spectra for complex networks}.
\newblock Cambridge University Press, 2010.

\bibitem{visintin1984preisach}
A.~Visintin.
\newblock On the {P}reisach model for hysteresis.
\newblock {\em Nonlinear Analysis: Theory, Methods \& Applications},
  8(9):977--996, 1984.

\bibitem{visintin2013differential}
A.~Visintin.
\newblock {\em Differential models of hysteresis}, volume 111.
\newblock Springer Science \& Business Media, 2013.

\bibitem{xiuqing1992spectral}
H.~Xiuqing and L.~Youyan.
\newblock Spectral structure and gap-labeling properties for a new class of
  one-dimensional quasilattices.
\newblock {\em Chinese Physics Letters}, 9(11):609, 1992.

\end{thebibliography}

\end{document}